\documentclass[preprint,showpacs,preprintnumbers,amsmath,amssymb,nofootinbib]{revtex4-2}

\setcitestyle{super}

\usepackage{graphicx}
\usepackage{braket}
\usepackage{natbib}
\usepackage{siunitx}
\usepackage{rotating}
\usepackage{amsmath,amssymb}%
\usepackage{amsfonts}
\usepackage{amssymb}
\usepackage{mathrsfs}
\usepackage{color}
\usepackage{bm}

\def\cblue{\color{black}}
\def\cred{\color{black}}

\begin{document}
\newcommand\blfootnote[1]{%
	\begingroup
	\renewcommand\thefootnote{}\footnote{#1}%
	\addtocounter{footnote}{-1}%
	\endgroup}

 \newcommand{\FigCap}[1]{\textbf{#1}}	

\title{Evidence for a finite-momentum Cooper pair in tricolor  $d$-wave  superconducting superlattices}

\author{T.\,Asaba$^{1,*}$}
\author{M.\,Naritsuka$^2$} 
\author{H.\,Asaeda$^1$} 
\author{Y.\,Kosuge$^1$}
\author{S.\,Ikemori$^1$}
\author{S.\,Suetsugu$^1$} 
\author{Y.\,Kasahara$^1$} 
\author{Y.\,Kohsaka$^1$} 
\author{T.\,Terashima$^1$} 
\author{A.\,Daido$^1$} 
\author{Y.\,Yanase$^1$} 
\author{Y.\,Matsuda$^{1,*}$}

\affiliation{$^1$Department of Physics, Kyoto University, Kyoto 606-8502 Japan}
\affiliation{$^2$RIKEN Center for Emergent Matter Science, Wako, Saitama 351-0198, Japan}

\maketitle

{\bf Fermionic superfluidity with a nontrivial Cooper-pairing, beyond the conventional Bardeen-Cooper-Schrieffer state, is a captivating field of study in quantum many-body systems. In particular, the search for superconducting states with finite-momentum pairs has long been a challenge, but establishing its existence has long suffered from the lack of an appropriate probe to reveal its momentum.  Recently, it has been proposed that the nonreciprocal {\cred electron} transport is the most {\cred powerful} probe for the finite-momentum pairs, {\cred because it directly couples} to the supercurrents. Here we reveal such a pairing state by the non-reciprocal  transport on tricolor superlattices with strong spin-orbit coupling combined with broken inversion-symmetry consisting of atomically thin $d$-wave superconductor CeCoIn$_5$.   We find that while the second-harmonic resistance exhibits a distinct dip anomaly at the low-temperature ($T$)/high-magnetic field ($H$) corner in the $HT$-plane for ${\bm H}$ applied to the antinodal direction of the $d$-wave gap, such an anomaly is absent for ${\bm H}$ along the nodal direction.  By meticulously isolating extrinsic effects due to vortex dynamics, we reveal the presence of a non-reciprocal response originating from intrinsic superconducting properties characterized by finite-momentum pairs. We attribute the high-field state to the helical superconducting state, wherein the phase of the order parameter is spontaneously spatially modulated.}

 A fundamental assumption of the Bardeen-Cooper-Schrieffer (BCS) theory of superconductivity is that two electrons form a Cooper pair with zero center-of-mass momentum. Realizing exotic superconducting states with finite-momentum pairs that violate this assumption has been a long-sought goal in condensed matter physics.    Such a superconducting state is an enticing theoretical possibility but has proven a severe experimental challenge. This is not only because the conditions under which such a superconducting state can be formed are rather stringent but also because smoking-gun experiments to confirm its existence have still been lacking.  

Helical superconductivity, in which the amplitude of the superconducting order parameter is constant but  its phase is spontaneously and spatially modulated, has been proposed as a prominent example of such a finite-momentum pair state~\cite{kaur2005helical,agterberg2007magnetic,akbari2022fermi,yuan2021topological}. The realization of helical superconductivity requires a strong Rashba effect that appears as a combined consequence of significant spin-orbit interaction (SOI) and spatial inversion symmetry breaking. When the crystal structure lacks a center of inversion, the SOI may dramatically change the electronic properties, leading to nontrivial quantum states.    The key microscopic ingredient in understanding the physics of such non-centrosymmetric materials is the appearance of antisymmetric SOI of the single electron states. Asymmetry of the potential in the direction perpendicular to the two-dimensional (2D) plane $\nabla V\parallel (001)$  induces Rashba type SOI,  $\alpha_R{\bm g}({\bm k})\cdot {\bm \sigma}\propto({\bm k}\times \nabla V)\cdot {\bm \sigma}$, where $\alpha_R$ is the Rashba coupling, ${\bm k}$ is the wave number,  ${\bm g}=(-k_y,k_x,0)/k_F$ with $k_F$ the Fermi wave number, and ${\bm \sigma}$ is the Pauli matrix \cite{bychkov1984properties}.  Rashba SOI splits the Fermi surface into two sheets with different spin structures. The energy splitting is given by $\alpha_R$, and the spin direction is tilted into the plane, rotating clockwise on one sheet and anticlockwise on the other.

The Rashba SOI has profound consequences on the superconducting states~\cite{smidman2017superconductivity,fischer2022superconductivity}. For example, parity is generally no longer a good quantum number, leading to exotic states with a mixture of spin-singlet and spin-triplet components. When the Rashba splitting becomes sufficiently larger than the superconducting gap energy,  it has been theoretically proposed that an even more fascinating superconducting state may emerge in 2D superconductors by applying strong parallel magnetic fields; a conventional BCS state with zero-momentum pairs $({\bm k}\uparrow,-{\bm k}\downarrow)$ formed within spin-textured Fermi surfaces (Fig.\,1a)  changes into a superconducting state with finite-momentum pairs formed within each spin nondegenerate Fermi surface~\cite{kaur2005helical,agterberg2007magnetic,akbari2022fermi,yuan2021topological}  (Fig.\,1b).   Such a superconducting state appears as a result of the shift of the Rashba-split Fermi surfaces by external parallel fields. When the magnetic field is applied parallel to $\hat{\bm x}$ axis (${\bm H}=H\hat{\bm x}$), the center of the two Fermi surfaces with different spin helicity are shifted along $\hat{\bm y}$ axis in opposite directions.   This state, referred to as a helical superconducting state,  is characterized by the formation of  Cooper pairs $({\bm k}+{\bm q_R}\uparrow,-{\bm k}+{\bm q_R} \downarrow)$, where  ${\bm q}_R=\mu_BH\hat{\bm y}/\sqrt{\alpha_R^2+\frac{2E_F}{m}}$ with Bohr magneton $\mu_B$, Fermi energy $E_F$ and quasiparticle mass $m$.  This pair formation leads to 
a state in which the magnitude of the superconducting order parameter is constant, while its phase rotates in space with period $\pi/|{\bm q}_R|$ as $\Delta({\bm r})$ =$\Delta_0e^{2i{\bm q_R}\cdot{\bm r}}$. 

 We note  that the helical state is essentially different from the Fulde-Ferrell (FF) and Larkin-Ovchinnikov (LO) states, in which finite-momentum Cooper pairs are formed between sections of the Zeeman-split Fermi surfaces~\cite{fulde1964superconductivity,larkin1965inhomogeneous} (Fig.\,1c).   A potential FF or LO state has been reported in several candidate materials, by showing a phase transition inside the superconducting state~\cite{matsuda2007fulde,wosnitza2018fflo} {\cred through the measurements of} magnetization~\cite{radovan2003magnetic}, specific heat~\cite{lortz2007calorimetric, bianchi2003possible, agosta2017calorimetric}, nuclear magnetic resonance~\cite{kakuyanagi2005texture, mayaffre2014evidence, kinjo2022superconducting}, thermal conductivity~\cite{kasahara2020evidence}, ultrasound~\cite{watanabe2004high,imajo2022emergent}, and scanning tunneling microscope~\cite{kasahara2021quasiparticle}.     
{\cred In the FF state,  the finite momentum Cooper pairs lead to the phase modulation of the superconducting order parameter,  which is difficult to detect directly.  In the LO state,  the spatial modulation of the superconducting order parameter due to such pairs gives rise to periodic nodal planes in the crystal.   However, it should be emphasized that no direct evidence showing such periodic nodes has been reported so far.  
	This is mainly due to the inherent challenge in directly measuring the momentum of Cooper pairs within a superconducting state, calling for a novel probe to investigate the Cooper pair momentum.}

Very recently, it has been theoretically proposed that superconducting states with finite-momentum Cooper pairs exhibit a current-direction dependent critical current, namely the superconducting diode effect ~\cite{yuan2022supercurrent,daido2022intrinsic,he2022phenomenological,ilic2022theory}.  This diode effect appears due to the non-reciprocal nature of the pair momentum-dependence of the free energy.     Notably, the diode effect is significantly enhanced upon entering the helical superconducting state both in $s$-wave \cite{daido2022intrinsic,ilic2022theory} and $d$-wave superconductors \cite{daido2022superconducting}. The enhancement naturally leads to characteristic behaviors of non-reciprocal electron transport (NRET) in general. Therefore, measurements of the NRET provide a powerful tool for revealing the helical superconductivity.
The resistance of a 2D  film can be described as $R=R_0(1+\gamma\mu_0{\bm H} \times \hat{\bm z} \cdot {\bm I})$, where $R_0$ and ${\bm I}$ are the resistance in the zero-current limit and an electric current, respectively. The coefficient $\gamma$  gives rise to different resistance for rightward and leftward electric currents and can be finite in non-centrosymmetric materials.    Unless the resistive transition in magnetic fields is very sharp due to strong pinning,  the NRET response can be obtained by measuring the second harmonic resistance $R_{2\omega}$.  {\cred The comparison between $R_{2\omega}$ at low frequencies in the DC limit and the differential in the critical current has been well-documented across various systems, and the general consensus is that if one is finite, the other will also be finite~\cite{ando2020observation,wu2022field}. }

 Non-reciprocal electron transport (NRET) has been studied in several superconductors~\cite{ando2020observation,itahashi2020nonreciprocal,zhang2020nonreciprocal}. However, it remains an arduous task to discern whether the observed NRET response stems from intrinsic superconducting phenomena, such as exotic pairing states that contain finite-momentum Cooper pairs. This is because the NRET response can also arise from extrinsic effects such as asymmetric vortex pinning at the edge, surface, and interface, the ratchet effect of the pinning center, and geometry-dependent Meissner shielding effects~\cite{lee1999reducing,olson2001collective,de2006controlled,reichhardt2015reversible}. To overcome this challenge, we fabricated tricolor Kondo superlattices comprised of atomically thin CeCoIn$_5$ layers and meticulously isolated the intrinsic superconducting effects, carefully eliminating the extrinsic effects.

CeCoIn$_5$  is a well-known heavy-fermion superconductor with the highest bulk $T_c$ of 2.3\,K,  in which  $d_{x^2-y^2}$ superconducting gap symmetry is well established~\cite{izawa2001angular,allan2013imaging}. The bulk CeCoIn$_5$ possesses the inversion center. Then, fabricating tricolor superlattices with an asymmetric $\cdots A/B/C/A/B/C\cdots$ arrangement, in which non-superconducting metals sandwich CeCoIn$_5$ with atomic layer thickness, we can introduce the global inversion symmetry breaking (Fig.\,2a)~\cite{shimozawa2014controllable,shimozawa2016kondo,naritsuka2017emergent,naritsuka2021controlling}. {\cred Given that this superlattice comprises three distinct materials, it will be designated as "tricolor" henceforth.}
This tricolor system provides an ideal platform for revealing the helical superconducting state for the following reasons. First, Ce atoms have a large SOI, and the condition that the Rashba-SOI well exceeds the superconducting gap has been confirmed in various superlattices of CeCoIn$_5$, including the present tricolor superlattice,  by the highly enhanced  upper critical field  from Pauli limited critical fields in bulk (see SI and \cite{naritsuka2021controlling}).  Second, Cooper pairs can be confined in atomic CeCoIn$_5$ layers, forming 2D superconductivity \cite{naritsuka2017emergent}.  Third is the strong electron correlation effect in CeCoIn$_5$.   It has been theoretically pointed out that the correlation further strengthens the effect of Rashba SOI~\cite{fujimoto2007electron}. Furthermore, the suppression of the orbital pair-breaking effects promotes the appearance of helical superconducting phases. These features make the CeCoIn$_5$ superlattice system unique and suitable for realizing helical superconductivity compared to weakly correlated systems. Finally, $d$-wave superconductors are expected to respond differently to in-plane magnetic fields directed for the nodal and anti-nodal directions, possibly allowing the intrinsic NRET to be extracted by changing the field direction.     
 
The tricolor superlattices with $c$-axis-oriented structure are epitaxially grown on MgF$_2$ substrate using the molecular-beam-epitaxy technique. The 3-unit-cell-thick (3-UCT) YbCoIn$_5$, 8-UCT CeCoIn$_5$ and 3-UCT YbRhIn$_5$ are grown alternatively, where YbCoIn$_5$ and YbRhIn$_5$ are conventional non-superconducting metals (Fig.\,2a). In these tricolor superlattices, all layers are not mirror planes, and broken inversion symmetry can be introduced along the stacking direction.   
{\cred Given the necessity for a precise in-plane application of the magnetic field in this study, we employed the 8-UCT CeCoIn$_5$ tricolor superlattice sample previously characterized in ref.~\cite{naritsuka2017emergent}. This referenced work extensively investigated the temperature and angular dependence of the upper critical field for the sample. Moreover, the presence of the strong Rashba SOI is confirmed by the suppressed Pauli limit~\cite{naritsuka2017emergent}.  For the sake of achieving a high current density and ensuring meticulous control over the current orientation, the sample underwent patterning utilizing a focused ion beam (FIB) as depicted in Fig.\,2b.}
We note that both the $T_c$ and upper critical field of this sample changed only slightly before and after FIB patterning (See Fig.\,S1). The sample becomes superconducting at $T_c$\,=\,0.83\,K defined as the temperature at which the dc resistance $R_{dc}$ drops to 50\% of its normal state value at the onset (Fig.\,2c).     Non-reciprocal transport measurements are carried out by the standard lock-in technique (see Materials and Methods in SI).  The $R_{2\omega}$ curves are anti-symmetrized with respect to a magnetic field. The misalignment of ${\bm H}$ from the ab plane is less than 0.05$^{\circ}$. 

Figure\,3 depicts $R_{2\omega}$ normalized by normal state resistance $R_n$ when both in-plane ${\bm H}$ and ${\bm I}$ are applied to nodal  (${\bm H}\parallel [110]$, ${\bm I}\parallel [1\bar{1}0]$) and antinodal (${\bm H}\parallel [100]$, ${\bm I}\parallel [010]$) directions (see the inset).   For both configurations,  finite $R_{2\omega}$  is detected only in the superconducting states, demonstrating that $R_{2\omega}$  originates from the Cooper pairs. Therefore,  despite the broadening of resistive transition that the inhomogeneity may cause, only the superconducting response of $R_{2\omega}$ can be extracted.  For the nodal configuration,  $R_{2\omega}$ increases with $H$ at low temperatures, peaking at $\mu_0H\sim 6$\,T and disappearing at high fields.  It should be noted that such a single-peak structure as a function of $H$ in the superconducting state has also been observed in NbSe$_2$~\cite{zhang2020nonreciprocal} and ion-gated SrTiO$_3$~\cite{itahashi2020nonreciprocal}. It was found that such a structure can be explained by the vortex motion. On the other hand, for the antinodal configuration, while a similar peak is observed at high temperatures, the peak is suppressed around $\mu_0 H\sim$ 5\,T at low temperatures,  exhibiting a distinct dip anomaly.

  {\cred 	Nonreciprocal responses can manifest even in the normal state in the presence of inversion symmetry breaking. However, {\cblue no discernible nonreciprocal response is observed in the normal state of the present superlattices.  Therefore, }the observed response can be attributed to the superconducting properties.} There are two possible sources for the  NRET response in the superconducting state.  One is extrinsic origin such as Meissner screening current and vortex motion, and the other is intrinsic origin due to the exotic superconducting state with finite momentum pairing.   We here discuss extrinsic origins.  The direction-dependent critical current can be  induced by the combination of the Meissner screening  effect and the asymmetric vortex surface barrier arising from the sample edges~\cite{hou2022ubiquitous}.   However, since such an effect is important only at a very low field around the lower critical field, it is negligibly small in the present field range{\cred , which significantly exceeds the Pauli limit.} 
  
 Another extrinsic origin is the asymmetric vortex motion.  In the present superlattices, the elliptical vortices may be formed within the CeCoIn$_5$ layer in a parallel field because the thickness of the  CeCoIn$_5$ layers is comparable to the $c$-axis coherence length.     When both ${\bm H}$ and ${\bm I}(\perp {\bm H})$ are applied in-plane, the vortices move in and out across the interfaces.   If there is an asymmetric vortex pinning potential at the interface of the different materials, NRET may occur.  In the tricolor superlattices,  different vortex thread pinning potentials on either side of the superconductor interface may induce NRETs of different amplitudes,  leading to an asymmetric motion that can generate a net NRET signal.  In fact, we measured the NRET response in bicolor $\cdots A/B/A/B \cdots$ stacking superlattices with canceling contributions on both sides of the interface and observed no NRET response (Fig. S5). This  supports the presence of the NRET response arising from vortex motion perpendicular to the layers.
 To rule out the possibility of pancake vortices perpendicular to the layer created by small but finite misalignment of ${\bm H}$ out of the 2D plane inducing the NRET effect, we measured the NRET by applying ${\bm H}$ tilted about 4$^{\circ}$ from the $ab$ plane and found no such effect in the tricolor superlattices (See Fig.\,S4). Additionally, no NRET effect was observed in the Lorentz force-free geometry, ${\bm H}\parallel {\bm I}$.  These results indicate that the extrinsic NRET response, if present, arises from asymmetric vortex motion perpendicular to the layers.

The present tricolor superlattices with tetragonal crystal symmetry have no twin boundaries. It is then unlikely that the vortex motion perpendicular to the 2D plane depends on the in-plane ${\bm H}$ and ${\bm I}(\perp {\bm H})$ directions.    To separate the intrinsic contribution from the extrinsic one, we take the difference between two configurations, as represented by the gray area in Fig.\,3.   Notably, except for the gray area at low temperatures, $R_{2\omega}$ from the  two configurations nearly overlaps.  
  This indicates that $R_{2\omega}$ for both configurations is dominated by extrinsic vortex motion except for the regime where $R_{2\omega}$  exhibits a dip anomaly at low temperatures around  $\mu_0H\sim 5$\,T for the antinodal configuration.   Therefore,   the dip anomaly when both ${\bm H}$ and ${\bm I}$ are applied to antinodal directions is attributed to an intrinsic origin arising from the Cooper pairs superimposed on the extrinsic vortex contribution. To obtain further information on the origin of the dip, $R_{2\omega}$ was measured at different relative angles of ${\bm H}$ and ${\bm I}$;  ${\bm H}\parallel $[100] and $I\parallel $[1$\bar{1}$0], and ${\bm H}\parallel $[110] and ${\bm I}\parallel $[010] (see SI for details). The results show that the appearance of the dip anomaly is determined by field direction,  not by the current direction, implying that the dip anomaly is related to the superconducting gap structure.

The NRET response provides pivotal information on the superconducting phase diagram of the tricolor superlattice displayed in Fig.\,4. The solid line in Fig.\,4 represents upper critical field  for ${\bm H}\parallel [100]$, $H_{c2\parallel}$,  determined by $H$ where $R_{dc}$ reaches 50\% of $R_n$. We find that $H_{c2\parallel}$ line for ${\bm H}$ applied nodal direction well coincides with that for anti-nodal direction, indicating a similar $HT$-phase diagram (Fig\,S1).  The upper right (colored in light brown) area in Fig.\,4 represents the normal state.   In the superconducting state below $H_{c2\parallel}$, the difference  of  $R_{2\omega}$ between two configurations,   $\Delta R_{2\omega}\equiv R_{2\omega}({\bm H}||[110],{\bm I}||[1\bar{1}0])-R_{2\omega}({\bm H}||[100],{\bm I}||[010])$,  normalized by $R_n$, is plotted in color; the gray area displayed in Fig.\,3 corresponds to $\Delta R_{2\omega}/R_n$.    In the light blue area at low fields in Fig.\,4, while finite $R_{2\omega}$ is observed for both configurations due to extrinsic contributions from vortex motion,   $\Delta R_{2\omega}$ is negligibly small. In the red area at high fields,   the finite $\Delta R_{2\omega}$ appears due to the intrinsic contribution originating from Cooper pairs. Note that as shown by arrows in Fig.\,3 indicating $H_{c2\parallel}$ determined by $R_{dc}$, $\Delta R_{2\omega}$ vanishes at around $H_{c2\parallel}$, while small but finite $R_{2\omega}$ remains at $H_{c2\parallel}$, likely due to the superconducting fluctuation effect and inhomogeneity.

{\cred No NRET is observed in the bicolor superlattice (Fig. S5).   In this system, as discussed above, the response arising from the vortex motion is canceled out.  {\cblue The fact that the bicolor superlattice preserves global inversion symmetry leads us to conclude that} the emergence of  NRET difference in the tricolor lattice, $\Delta R_{2\omega}$,  is an intrinsic phenomena arising from the Rashba SOI.    } The intrinsic NRET emerges as a direct consequence of the state with finite-momentum  pairs and such an effect is negligibly small in the BCS state.   Therefore, the results of Fig.\,4 provide evidence for the appearance of a high-field superconducting state at the low-$T$/high-$H$ corner, distinct from the low-field BCS state.    Although the anomalous upturn behavior of $H_{c2\parallel}$ at low temperatures has been suggested in the previous study~\cite{naritsuka2017emergent}, the superconducting state at high fields had remained an unresolved issue, including the possible existence of a new phase. {\cred We note that we can rule out the possibility that the observed nonreciprocal phenomena are tied to the so-called $Q$-phase~\cite{kenzelmann2010evidence}{\cblue, in which the superconductivity may be intertwined with magnetic order, in bulk CeCoIn$_5$ } for the following reasons. Firstly, in the basic Drude model, nonreciprocal transport is independent of spin. Then, the primary effect of the $Q$-phase on nonreciprocal transport is the Brillouin zone folding, but this has a negligible effect. In addition, in the $Q$-phase, where spatial modulations {\cblue of order parameter} appear, electron scattering should increase, resulting in the suppression of the nonreciprocal response. However, our observations indicate the opposite in the current case. Furthermore, unlike the FFLO states, Cooper pairs in the $Q$-phase do not carry a finite momentum. Hence, even when the inversion symmetry is broken, the momentum of the Cooper pair remains unchanged.  } 
 
There are two scenarios for finite-momentum pairing states other than the helical superconducting state: the FF and LO states~\cite{matsuda2007fulde} (Fig.\,1c), where pairing between sections of the Zeeman split Fermi surfaces results in Cooper-pairs $({\bm k}+{\bm q}_Z\uparrow, -{\bm k}+{\bm q}_Z\downarrow)$ with  momentum ${\bm q}_Z\approx\mu_B{\bm H}/\hbar v_F$ ($v_F$ is the Fermi velocity).  In the FF state,  the superconducting order parameter is described as $\Delta({\bm r})\propto \exp(2i{\bm q_Z}\cdot {\bm r})$ with constant amplitude and spatially varying phase, while in the LO state with $\Delta({\bm r})\propto \cos(2{\bm q_Z}\cdot {\bm r})$,  the amplitude oscillates in space.   However, it should be stressed that we can discard the possibility of both FF and LO states as the origin of the intrinsic NRET phenomena.  This is because the energy of the Rashba spin splitting is overwhelmingly larger than the superconducting gap energy in the present tricolor superlattices, as demonstrated in~\cite{naritsuka2017emergent} (see Fig.\,S2).  In this situation, FF- and LO-type pair formation cannot occur.  In addition,  since the LO state contains ${\bm q_Z}$ and ${-\bm q_Z}$, NRET should be canceled out. {\cred In the absence of inversion symmetry, the LO phase can be characterized by a general order parameter, formulated $a e^{2iq_{Zr}} + b e^{-2iq^{\prime}_{Zr}}, a \neq b$. This phase, as defined by the order parameter, is commonly refered to as the 'stripe phase'. However,  theoretical predictions suggest that the stripe phase only emerges within a narrow region at low temperatures in the $HT$-phase diagram \cite{agterberg2007magnetic}, whereas helical superconductivity manifests within a more expansive phase region surrounding it. Considering this, it seems plausible that our experimental observations represent helical superconductivity. It may be possible, however, to observe a stripe phase at even lower temperatures.}  Based on these results, we conclude that the high field regime indicated by the red color in Fig.\,4 represents the helical superconducting state, and the low field regime by light blue corresponds to the BCS state.

The strong field-orientation dependence of intrinsic NRET likely appears as a result of the direction-dependent Doppler shift of the quasiparticles in $d$-wave superconductors. When ${\bm H}$ is applied parallel to the nodal direction, quasiparticles around the nodes perpendicular to the magnetic field are excited. When the current is applied parallel to these nodes, the system exhibits more metallic behavior compared with that for the antinodal direction. Although such a simple interpretation should be scrutinized,  the present results also point toward the importance of nodal structure for the direction-dependent NRET.   We note that the finite-momentum pairing state has been suggested in the pair-density-wave (PDW) state in the pseudogap phase of cuprates by scanning tunneling microscope measurements~\cite{hamidian2016detection}.  Therefore, it is highly intriguing to apply the present direction dependent NRET to the putative PDW state.  

The NRET effect arising from the intrinsic superconducting response observed in the tricolor $d$-wave superconducting superlattice with strong Rashba interaction provides evidence for the emergence of  a superconducting state with finite-momentum Cooper pairs at high fields, most likely a helical superconducting state.  Such a unique state provides a  platform to investigate the novel fermionic superfluid systems beyond the BCS pairing states.

\bibliography{TricolorNotes}

\newpage
\begin{figure}[t]
	\centering
	\includegraphics[width=\linewidth]{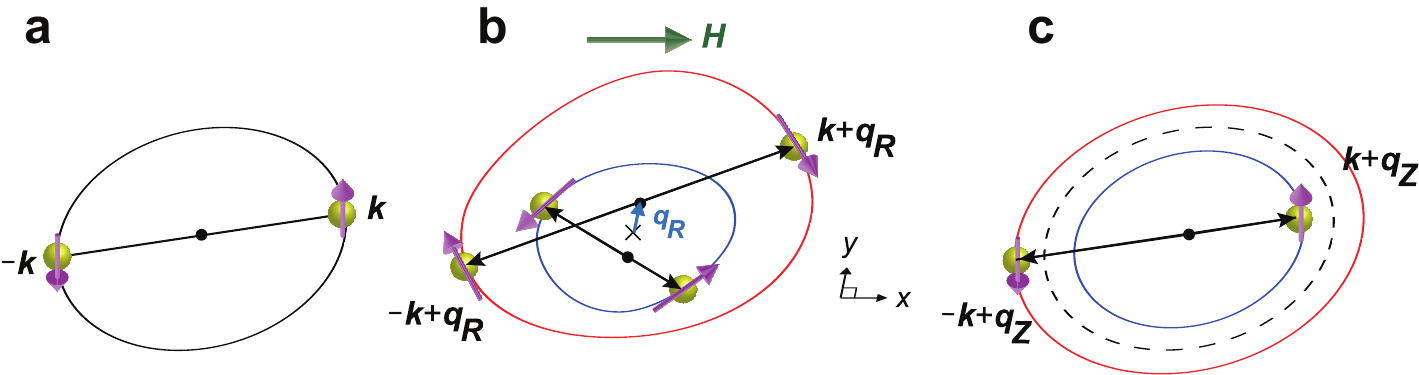}
	\caption{\label{Fig:basic} {\bf  Schematic of the various types of Cooper pairings.} {\bf a,} Conventional BCS pairing  state.   Zero momentum pairing with $({\bm k}\uparrow, -{\bm k}\downarrow)$ occurs between electrons in states with opposite momentum and opposite spins.   {\bf b,}   Helical superconducting state. 	Arrows on the Rashba-split Fermi surfaces indicate spins.  ${\bm H}$ parallel to $\hat{\bm x}$ axis shifts the center of the small and large Fermi surfaces by ${\bm q}_R$ along $+y$ and $-y$ directions respectively.  Pairs are formed within each Rashba split Fermi surface between the states of $({\bm k}+{\bm q}_R\uparrow, {\bm -k}+{\bm q}_R\downarrow)$, leading a gap function with modulation of phase $\Delta({\bm r})\propto \exp(2i{\bm q_R}\cdot {\bm r})$.    Cooper-pairs have  finite center-of-mass momentum ${\bm q}_R$.  {\bf c,} FF and LO pairing states.   Pairing with $({\bm k}+{\bm q_Z}\uparrow, -{\bm k}+{\bm q_Z}\downarrow)$ occurs between sections of the Zeeman split Fermi surfaces, where  ${\bm q}_Z\approx 2\mu_B{\bm H}/\hbar v_F$.  Cooper-pairs have  finite centre-of-mass momentum ${\bm q}_Z$.  In FF state, the order parameter varies as $\Delta({\bm r})\propto \exp(2i{\bm q_Z}\cdot {\bm r})$, while in the LO state, it varies as $\Delta({\bm r})\propto \cos(2{\bm q_Z}\cdot {\bm r})$.}
\end{figure}
\newpage
\begin{figure}[t]
	\centering
	\includegraphics[width=\linewidth]{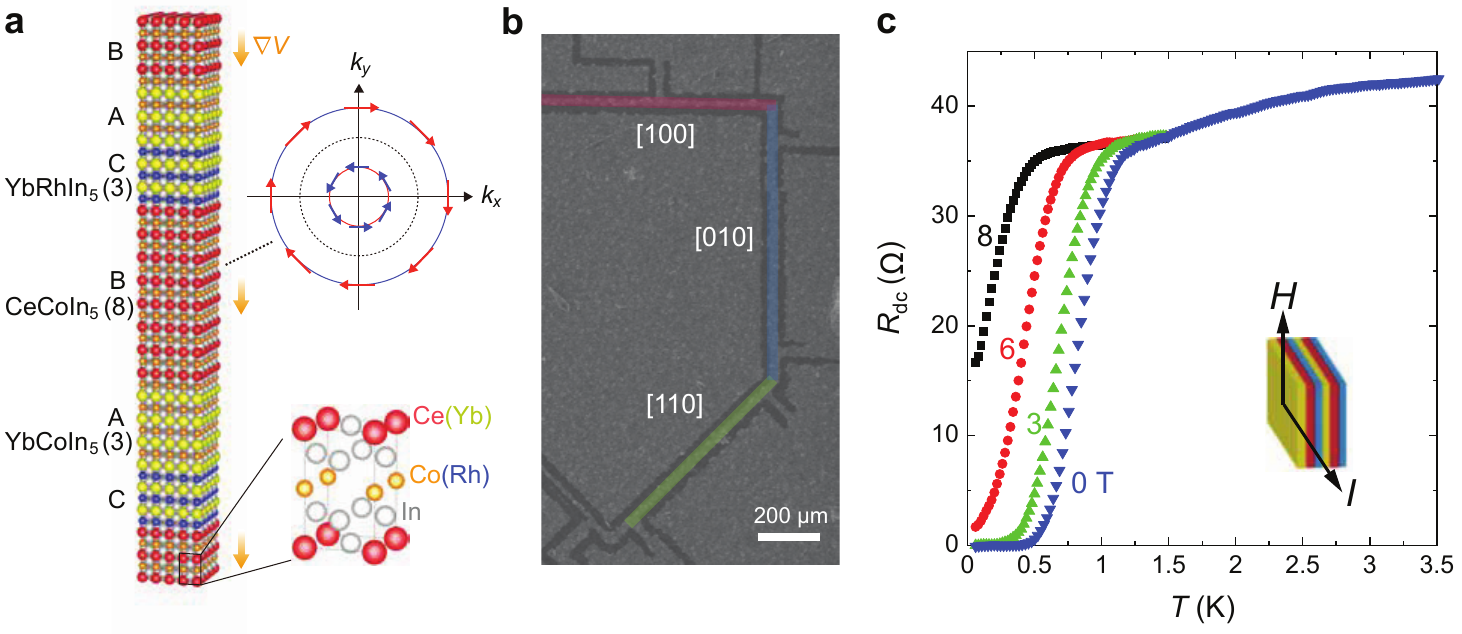}
	\caption{\label{Fig:basic} {\bf Tricolor $d$-wave superconducting superlattices } {\bf a,}Schematic representation of noncentrosymmetric tricolor Kondo superlattices with $\cdots A/B/C/ A/B/C\cdots$ structure.  The sequence of YbCoIn$_5$(3)/CeCoIn$_5$(8)/YbRhIn$_5$(3) is stacked repeatedly for 30 times, so that the total thickness is about 300\,nm.  The orange arrows represent the asymmetric potential gradient $\nabla V$, which gives rise to Rashba splitting of the Fermi surface with different spin structure.  The crystal structure of Ce(Yb)Co(Rh)In$_5$ is also illustrated.   {\bf b,}  The scanning electron microscopy image of a tricolor superlattice patterned by focused-ion-beam (FIB). The black line regime corresponds to the area cut by FIB. Red, blue and green lines indicate the current path along [100], [010], and [110], respectively. The width of the current path is $20\pm2$\,$\mu$m. {\bf c,} The temperature dependence of the dc resistance $R_{dc}$ for ${\bm H}\parallel $[100] and  ${\bm I}\parallel $ [010].   } 
\end{figure}
\begin{figure}[t]
	\centering
	\includegraphics[width=0.6\linewidth]{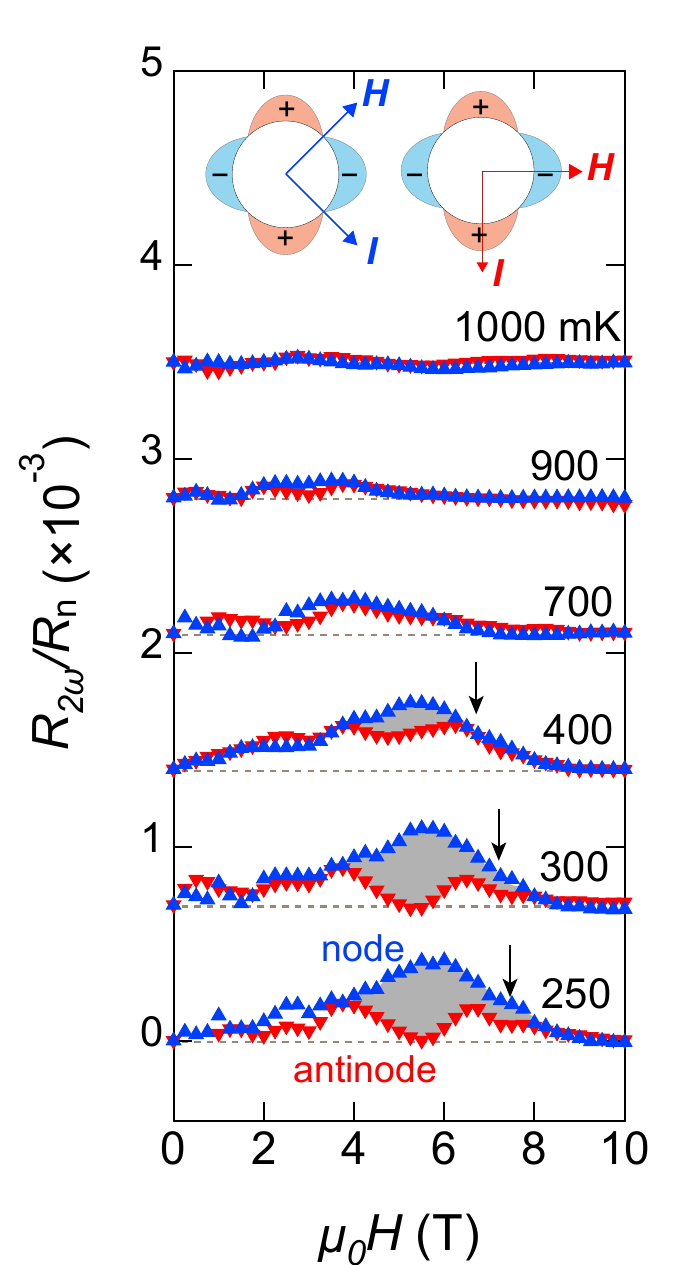}
	\caption{\label{Fig:basic} {\bf Non-reciprocal electronic transport in the superconducting state. }  The field dependence of second harmonic resistance $R_{2\omega}$ normalized by $R_n$ for two configurations.  Blue upper triangles show  $R_{2\omega}/R_n$ for both ${\bm H}$ and ${\bm I}$  applied parallel to the $d$-wave nodal direction (${\bm H}\parallel $[110], ${\bm I}\parallel $[1$\bar{1}$0]), as illustrated by  left panel in the inset.   Red lower triangles show  $R_{2\omega}/R_n$  for  antinodal configuration (${\bm H}\parallel $[100], ${\bm I}\parallel $[010], right panel in the inset).  
The curves are vertically shifted for clarity. The dashed lines indicate the base lines. The gray area represents the difference between two configurations $\Delta R_{2\omega}/R_n$.  
	Arrows indicate $H_{c2\parallel}$ determined by $R_{dc}$. }
\end{figure}

\begin{figure}[t]
	\centering
	\includegraphics[width=\linewidth]{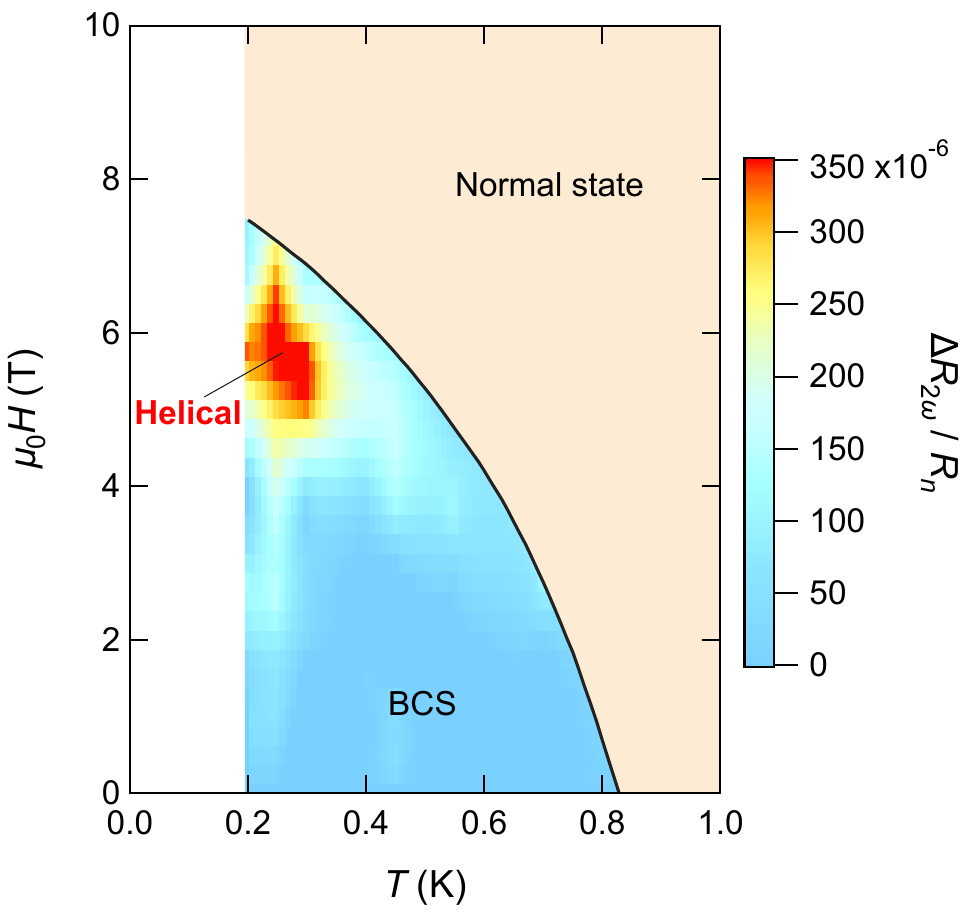}
	\caption{\label{Fig:basic} {\bf  Superconducting phase diagram of the tricolor superlattice determined by nonreciprocal electron transport properties.}  The solid line is $H_{c2\parallel}$ determined by  $R_{dc}$.  $H_{c2\parallel}$ line  for  nodal and anti-nodal directions well coincides each other.    The light brown area  represents the normal regime.   In the superconducting state, the difference  of  $R_{2\omega}/R_n$ between two configurations, $\Delta R_{2\omega}/R_n$,  is plotted in color.  The blue area at low fields represents the BCS regime, where $\Delta R_{2\omega}/R_n$ is negligibly small while finite $R_{2\omega}$ is observed for both configurations due to extrinsic contributions from vortex motion.    The red area at high fields corresponds to helical  superconducting state, where the finite $\Delta R_{2\omega}/R_n$  appears due to the intrinsic contribution originating from Cooper pairs.}
\end{figure}
 
\end{document}